\documentclass[12pt]{iopart}
\usepackage{iopams}
\usepackage{graphicx,float}
\usepackage{orcidlink}
\usepackage{color}
\newcommand{\RR}{{L}}
\newcommand{\be}{\begin{equation}}
\newcommand{\ee}{\end{equation}}
\newcommand{\bea}{\begin{eqnarray}}
\newcommand{\eea}{\end{eqnarray}}
\newcommand{\dSL}{\delta S^{\rm left}_\mathrm{FLT}}
\newcommand{\dSR}{\delta S^{\rm right}_\mathrm{FLT}}
\newcommand{\dSFLT}{\delta S_\mathrm{FLT}}
\newcommand{\sone}{0.743744(3)}
\newcommand{\myeps}{10^{-12}}
\usepackage{comment}

\eqnobysec
\def\1.2{\frac{1}{2}}

\begin{document}
\title[Kondo Screening Cloud Scaling]
{Kondo Screening Cloud Scaling: Impurity Entanglement and Magnetization}
\author{Erik S. S{\o}rensen\orcidlink{0000-0002-5956-1190}}
\address{Department of Physics \& Astronomy, McMaster University, Hamilton ON L8S 4M1, Canada.}
\ead{sorensen@mcmaster.ca}

\date{\today}
\begin{abstract}
The screening of an impurity spin in the Kondo model occurs over a characteristic length scale $\xi_K$, that defines the size of the Kondo screening cloud or ``mist". The presence of such a length scale in experimental and numerical results is rather subtle. 
A consistent way to show the presence of the screening cloud  is to demonstrate scaling in the spatial correlations depending on 
$r$, in terms of the single variable $r/\xi_K$ rather than depending on $r$ and $\xi_K$ separately.
Here we study the paradigmatic one channel Kondo model using a spin chain representation, with an impurity spin at one end of the chain
coupled with a strength $J_K'$. Using Fermi liquid
theory combined with numerical results, we obtain new high precision estimates of the non-universal terms in the entanglement entropy which leads
to a verification of the expected non-integer ground-state degeneracy, $g$. This then allows us 
to study the impurity contribution to the entanglement in detail.
If the impurity coupling $J_K'$ is varied, a precise determination of $\xi_K$
can then be obtained. The length scale, $\xi_K$, is then shown to characterize the scaling of both the uniform and alternating part of a measure of
the magnetization of part of an odd length chain with.
\end{abstract}

\maketitle

\section{Introduction}
The Kondo effect~\cite{Kondo1964,Hewson,KondoRMP} continues to yield surprising insights. 
In its simplest form, the Kondo effect describes the coupling of a single $S=1/2$
impurity spin to a bath of non-interacting conduction electrons.
In the absence of any coupling, the electrons are undisturbed by the impurity, 
while at very strong coupling, a single electron is trapped by the impurity, forming a singlet at the impurity site.
Intuitively, one might then expect that at intermediate coupling the low energy physics is again described by the formation of
a singlet state between the conduction electrons and the impurity spin, often described as a screening cloud~\cite{Affleck2010}.
Only a single electron is needed to form this singlet, but in this case the screening singlet could be macroscopic in size. 
The natural length scale associated with this phenomenon is $\xi_K=\hbar v/k_BT_K$
with $v$ the velocity of the low energy excitations and $T_K$ the Kondo temperature, and for $T_K\simeq 1K$, could extend into the $\mu m$
regime for typical values of $v_F$. It is important to realize that this screening cloud~\cite{Affleck2010} which forms the singlet with the impurity only involves a single electron even though it is of course a many-body state, and it is perhaps better described as a ``mist". The presence of the screening cloud is therefore a particular
subtle feature of the Kondo effect, since it involves spatial correlations around the impurity out to large distances involving a single electron.
It is natural to ask what it actually means to have a screening cloud. The best definition is that physical quantities should obey a scaling
form~\cite{sorensen1996} in $r/\xi_K$ instead of depending on $r$ and $\xi_K$ independently, this then implies variations over at least a distance
of $\xi_K$. In fact, it is by demonstrating such scaling in experimental data that the presence of the Kondo screening cloud has recently been detected in experiments on quantum dots coupled to a quasi-one-dimensional channel~\cite{Borzenets2020}.

A focus of particular current interest has been the entanglement between the impurity and the electrons~\cite{sorensen20071CK}. In the strong coupling limit where
a single electron is trapped at the impurity site into a singlet state, other electrons are not allowed onto the site since it would destroy the
singlet, and we do not expect the impurity to significantly alter the entanglement. However, for intermediate couplings, the impurity contribution
to the entanglement can be shown to again follow a scaling form~\cite{sorensen20071CK,sorensen2007QIE}. At finite temperatures, the negativity is a convenient measure of the impurity entanglement~\cite{Bayat2010} and the thermal decay of the negativity has been shown to obey a characteristic  power-law form~\cite{Lee2015,Kim2021}. More generally, the entanglement across conformal and topological defects has been investigated~\cite{Roy2022,Rogerson2022,Capizzi_2023} and the Kondo effect in spin liquids has been studied~\cite{Kolezhuk2006,Dhochak2010,Das2016,He2022,Lu2022}, but the corresponding entanglement has so far only received limited attention. A difficulty is here that non-universal terms appear in the entanglement entropy that are often not known with sufficient precision. Furthermore,
the impurity contribution to the entanglement is defined by subtracting the bulk contribution, which is often a challenging procedure.
Here we revisit the one channel Kondo model and
study the impurity entanglement in detail. With open boundary conditions a contribution arising from the non-integer ground-state degeneracy, $g$\cite{affleck91c}, can only be precisely determined if the impurity entanglement from open boundaries are taken into account. We then turn to the
magnetization, demonstrating similar scaling with $r/\xi_K$.

We shall here take the coupling between the impurity spin and the conduction electrons to be a delta-function.
The spherical symmetry of the coupling implies that the problem can be reduced to a one-dimensional model after an expansion in harmonics
since only the s-wave harmonic interacts with the impurity.
If one linearizes the dispersion relation around the Fermi momentum, $k_F$, one can then write
a one-dimensional effective Hamiltonian in terms of left and right moving Fermi fields that are functions of $(v_Ft+x)$ and $(v_Ft-x)$, respectively,
obeying the anti-commutation relation $\{\psi_{L/R}(x),\psi^\dagger_{L/R}(y)\} = 2\pi \delta (x-y)$:
\begin{equation}
H=\frac{v_F}{2\pi }\int_0^\infty dx\left[ i\psi_L^\dagger \frac{d}{dx}%
\psi_L-i\psi_R^\dagger \frac{d}{dx}\psi_R\right] 
+v_F\lambda_K 
\vec S_\mathrm{el}(0)\cdot \vec S_\mathrm{imp}.
\label{eq:H1D}
\end{equation}
Here, $x$ is the radial coordinate and the impurity is now situated at $x=0$ with the electrons confined to a semi-infinite line
and
\begin{equation}
\vec S_\mathrm{el}(0)=\psi^\dagger_L(0)\frac{\vec \sigma}{2}\psi_L(0)
\end{equation}
is the electron spin density at $x=0$ and the boundary condition  $\psi_L(x=0)=-\psi_R(x=0)$ is imposed at the origin. Here,
$\lambda_K$ is the coupling between the impurity spin and the electron spin density.
(See, for example, \cite{affleck90} for more details.) 
Since this model explicitly refer to electronic degrees of freedom,
we refer to this model as the free electron Kondo model (FEKM). 

A model closely to the one-dimensional FEKM consists of an open gap-less spin-$\1.2$ Heisenberg chain,
defined on a semi-infinite line, with one coupling at the
end of the chain weaker than the others,
described by the following  Hamiltonian.
\begin{equation}
H_\mathrm{1CK} =J_{K}^{\prime }\left( \vec{S}_{1}\cdot \vec{S}_{2}+J_{2}\vec{S}%
    _{1}\cdot \vec{S}_{3}\right) +
\sum_{r=2}^{\RR-1}\vec{S}_{r}\cdot \vec{S}_{r+1}+J_{2}\sum_{r=2}^{\RR-2}\vec{%
  S}_{r}\cdot \vec{S}_{r+2}.  \label{eq:H1CK}
\end{equation}
It turns out that a low energy field theory description of this spin chain model is essentially the same as
Eq.~(\ref{eq:H1D}), and one may therefore think of Eq.~(\ref{eq:H1CK}) as a spin chain 
Kondo model (SCKM)~\cite{sorensen20071CK,sorensen2007QIE,laflorencie2008kondo} equivalent to the one-channel Kondo model.
In fact, it turns out that in the limit $J_K'\to 0$, a very simple linear relation exists between $\lambda_K$ in Eq.~(\ref{eq:H1D}) and $J_K'$ in Eq.~(\ref{eq:H1CK}), $J_K'=1.381\lambda_K$~\cite{laflorencie2008kondo}.
For $J'_K$=1 this model is the well-known $J_1$-$J_2$ spin model that with $J_2$=0 is simply the antiferromagnetic
Heisenberg chain. However, we shall only be interested in a very specific value of $J_2$, corresponding to a critical point~\cite{Eggert1996}:
\begin{equation}
    J^c_2=0.24116
\end{equation}
At this value of $J_2$ a bulk marginally irrelevant coupling becomes exactly zero~\cite{eggert1992magnetic}, leading to the disappearances of logarithmic corrections, a significant advantage for numerical work. For $J_2>J_2^c$ the SCKM develops a gap and orders in a dimer state, and the correspondence to the FEKM no longer holds. Another advantage of the SCKM is the significant reduction in the on-site
degrees of freedom, which means that higher precision results can be obtained with fewer resources. 

As a measure of the entanglement, we mainly focus on the von Neumann entanglement entropy. For a subsystem of linear length
$x$ the entanglement entropy is defined in terms of the corresponding reduced density matrix, $\rho_x$,
\begin{equation}
S(x)=-\Tr\rho_x\log\rho_x,
\end{equation}
where $\rho_x$ is obtained from the ground-state wave-function. To study the entanglement entropy and the magnetization in the SCKM,  we use density matrix renormalization group techniques~\cite{White1992a,White1992b,White1993,Schollwock2005,Hallberg2006,Schollwock2011,itensor} (DMRG) on finite chains of length up to $L=400$, typically with $\epsilon=\myeps$.

We begin with an analysis of quantum impurity entanglement in the SCKM, building on previous studies ~\cite{sorensen1996, sorensen20071CK,sorensen2007QIE,laflorencie2008kondo,affleck2009Review}. Here we significantly extend and refine many of the results. All the results in this section are obtained using the SCKM, Eq.~(\ref{eq:H1CK}) at $J_2$=$J_{2c}$=$0.241167$. As mentioned above, at this value of $J_2$ 
corrections arising from the bulk marginal operator are absent, significantly improving the precision of the results.
In the subsequent sections, we then turn to a discussion of our results for the magnetization.

\section{Entanglement in the Periodic Chain, $s_1$}
\begin{figure}[!ht]
\hfill\includegraphics[width=14cm,clip]{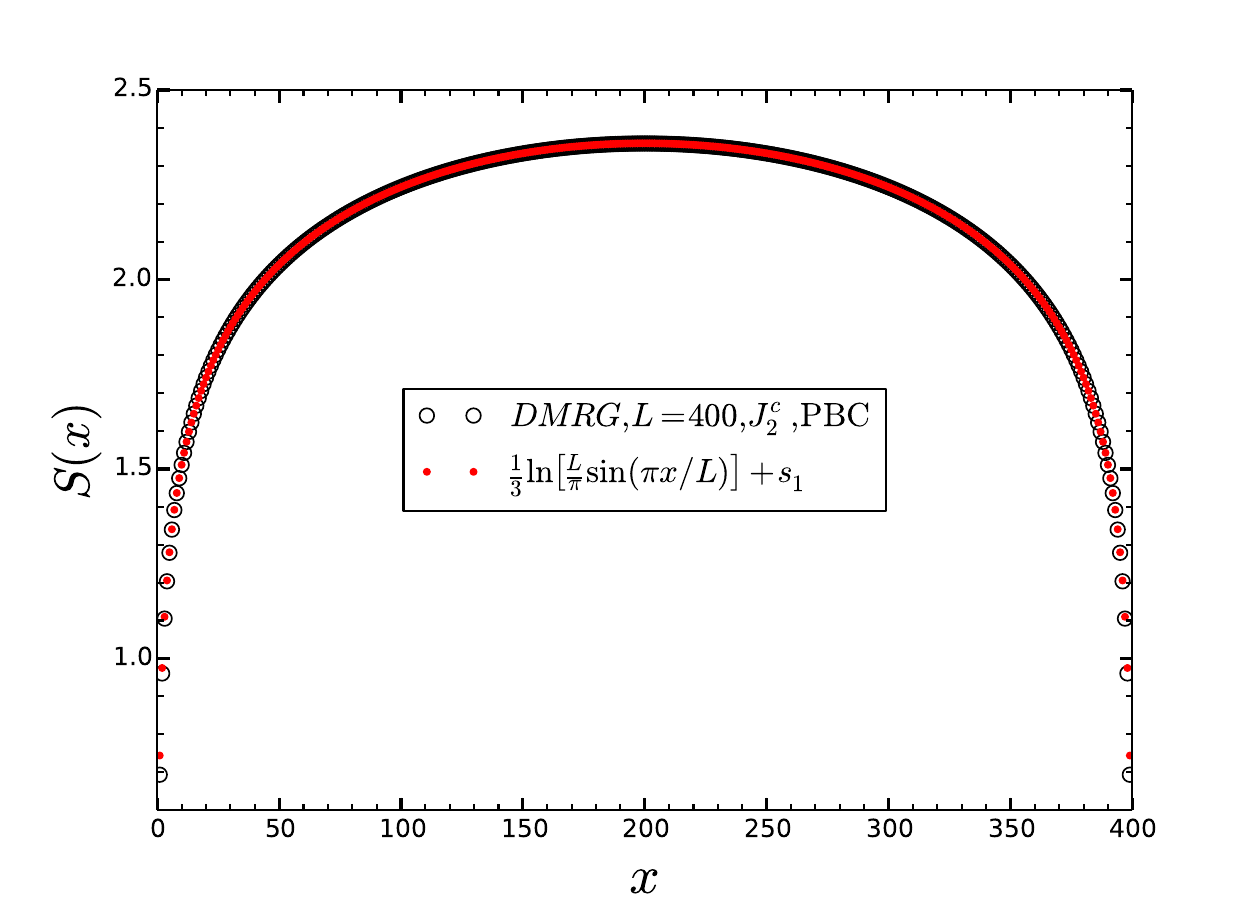}
\caption{The entanglement entropy, $S(x)$, as a function of $x$ for a L=400 site {\it periodic} chain with $J_2=J_2^c=0.241167$. The black circles indicate the numerical DMRG results while the solid red circles are 
a fit to the analytical form, Eq.~(\ref{eq:SPBC}), with $s_1=\sone$.
}
\label{fig:1CK_SPBC}
\end{figure}
First, we focus on the purely {\it periodic} chain with $J_2=J_2^c=0.241167$ since this will allow us to determine non-universal constants that are crucial for our subsequent analysis.
From analytical results~\cite{holzhey94,calabrese04} it is known that for a finite {\it periodic} system of $L$ sites, the entanglement entropy of a sub-system of size $x$ is:
\begin{equation}
S^\mathrm{pbc}(x)=\frac{c}{3}\ln\left[\frac{L}{\pi}\sin\left(\frac{\pi x}{L}\right)\right]+s_1,
\label{eq:SPBC}
\end{equation}
where $c$ is the central charge and $s_1$ is a non-universal constant. For the models we consider, we have $c=1$. Our DMRG results with $\epsilon=\myeps$ for $S(x)$ are shown in Fig.~\ref{fig:1CK_SPBC} for a {\it periodic} chain with
$J_2=J_2^c=0.241167$. With periodic boundary conditions (PBC) there is no boundary and $S(x)$ is smoothly varying as the size of the sub-system $x$ changes. The DMRG results can be fitted extremely well to the expected form, Eq.~(\ref{eq:SPBC}), from which we obtain at $J_2^c$:
\begin{equation}
s_1=\sone\ \ (J_2=J_2^c=0.241167).
\label{eq:s1}
\end{equation}
Note that, $s_1$ is non-universal and for instance depends on $J_2$. Hence, at the isotropic Heisenberg point with $J_2=0$, we find a slightly smaller value of $s_1\sim 0.73771(1)$. One may also note that $s_1$ the for the non-interacting $XX$ chain has been determined to be $s_1=0.726\ 067$~\cite{laflorencie2006}. The variation in $s_1$ between the different models is surprisingly small.

\section{Fermi Liquid Theory}
Consider now the uniform $S$=1/2 SCKM chain with open boundary conditions (OBC) of interest here.  The chain has length $L$ and we consider 
a region at one end of the open chain of length $x$.  
As shown in \cite{laflorencie2006}, one effect of the open boundary conditions is to induce an alternating term $S_A$, in the
von Neumann entanglement entropy, $S(x)$. The latter is therefore best view as composed of a uniform and alternating term
\begin{equation}
   S^\mathrm{obc}=S^\mathrm{obc}_U(x)+(-1)^xS^\mathrm{obc}_A(x) 
\end{equation}
with both $S_U$ and $S_A$ smoothly varying functions. If the bulk of the system is uniform, the alternating term  $S_A^\mathrm{obc}(x)$ decays to zero with $x$. However, recent work has shown that if the sub-system starts at the boundary and ends at an impurity, then the alternating term does {\it not} necessarily decay to zero~\cite{Schlomer2022}.
The bulk contribution to the uniform part of the von Neumann entanglement entropy, $S_U^\mathrm{obc}$, is given by~\cite{holzhey94,calabrese04,calabrese06,Zhou2006}:
\begin{equation}
S_U^\mathrm{bulk}(x)=\frac{c}{6}\ln \left[\frac{2L}{\pi}\sin\left(\frac{\pi x}{L}\right)\right]+\frac{s_1}{2}+\ln g.
\label{eq:Sbulk}
\end{equation}
In contrast to $s_1$ the constant $\ln g$ is a universal term arising from a non-integer ground-state degeneracy, $g$~\cite{affleck91c}. 
However, this is not the only contribution to $S^\mathrm{obc}_U$ since  the open ends of the chain gives rise to a boundary term, $\delta S_U$~\cite{sorensen20071CK,sorensen2007QIE}, so that
\begin{equation}
S_U^\mathrm{obc}(x)=S_U^\mathrm{bulk}(x)+\delta S_U(x).
\end{equation}
It is possible to describe the leading contributions to $\delta S_U$ using conformal field theory (CFT) methods based on Nozi\`eres local Fermi liquid theory (FLT)~\cite{nozieres,affleck90,affleck91a,affleck91b}.
The leading contributions, due to the two open ends, are caused by the leading irrelevant operator 
$\eta [T(0)+T(L)]$ where $T(x)$ is the energy density, $\propto (\partial_x\phi )^2$ in Abelian bosonization, and $\eta$ is  a coupling constant of order a lattice spacing. The first 
order correction due to $T(0)$, the boundary interaction at the left end of the system, was calculated in Ref.~\cite{sorensen20071CK}:
\begin{equation}
\dSL=
{\eta \over 12L}\left[ 1+\pi{L-x\over L}\cot (\pi x/L)\right]  .
\label{eq:dSleft}
\end{equation}
This result is 
valid in the FLT regime:
\begin{equation}
\xi_K \ll x \ll L
\label{eq:FLTreg}
\end{equation}
There is another correction due to $T(L)$, the boundary operator at the right end of the system.  This can be obtained by 
substituting $x\to L-x$, thereby obtaining:
\begin{equation}
\dSR=
{\eta \over 12L}\left[ 1-\pi{x\over L}\cot (\pi x/L)\right],
\label{eq:dSright}
\end{equation}
where we have used the standard relation $\cot(\pi-x)=-\cot(x)$.
Since the analysis in Refs.~\cite{sorensen20071CK,sorensen2007QIE} involved a subtractive procedure, approximately eliminating $\dSR$, this term was not included.
Following the convention adopted in~\cite{sorensen20071CK,laflorencie2008kondo} we identify $\eta$ with $\pi\xi_K$,
where $\xi_K$ can be identified with the size of the Kondo screening cloud.
With this normalization, the zero-temperature impurity susceptibility is $\chi_{imp}=1/(4T_K)$.
The total correction arising from both ends of the open chain is then:
\begin{equation} 
\dSFLT=
\dSL+\dSR={\pi\xi_K \over 12L}\left[ 2+\pi{L-2x\over L}\cot (\pi x/L)\right]  .
\label{eq:dSFLT}
\end{equation}
This is symmetric under $x\to L-x$ and is functionally similar to $1/[L\sin (\pi x/L)]$ up to an overall multiplicative constant and in
Ref.~\cite{laflorencie2006} the leading correction term to Eq.~(\ref{eq:Sbulk}) arising from the open boundary was therefore assumed to be of the form $1/(L\sin(\pi x/L))$ instead of the correct form, Eq.~(\ref{eq:dSFLT}).

If we consider a slightly modified coupling, $J_K^\prime$, of the spin at the left end of the chain, the 
boundary operator $T(0)$ will still appear but with 
a different coupling constant which we can write as $\pi \xi_K^\mathrm{left}$. 
If we denote the coupling on the right side as $\pi \xi_K^\mathrm{right}$, we obtain:
\begin{eqnarray}
S^\mathrm{obc}_U = S_U^\mathrm{bulk}+ \dSL+\dSR\nonumber\\
\dSL = 
{\pi\xi_K^{\rm left} \over 12L}\left[ 1+\pi{L-x\over L}\cot (\pi x/L)\right]\nonumber\\
\dSR  = 
{\pi\xi_K^{\rm right} \over 12L}\left[ 1-\pi{x\over L}\cot (\pi x/L)\right],
\label{eq:sb}
\end{eqnarray}
with $S_U^\mathrm{bulk}$ from Eq.~(\ref{eq:Sbulk}).
Note that in the limit $L\to\infty$ we have~\cite{sorensen20071CK}:
\begin{equation}
\dSL  \sim \pi\xi_K^{\rm left}/(12 x),
\label{eq:dSthermo}
\end{equation}
with a similar expression for $\dSR$. Hence, both $\dSL$ and $\dSR$ remain finite in the thermodynamic limit at fixed $x$.
On the other hand, they can be neglected if $x/L$ is kept fixed as $L\to\infty$.

\section{Impurity Entanglement in the Open Chain, $\ln g$}
\begin{figure}[!ht]
\hfill\includegraphics[width=14cm,clip]{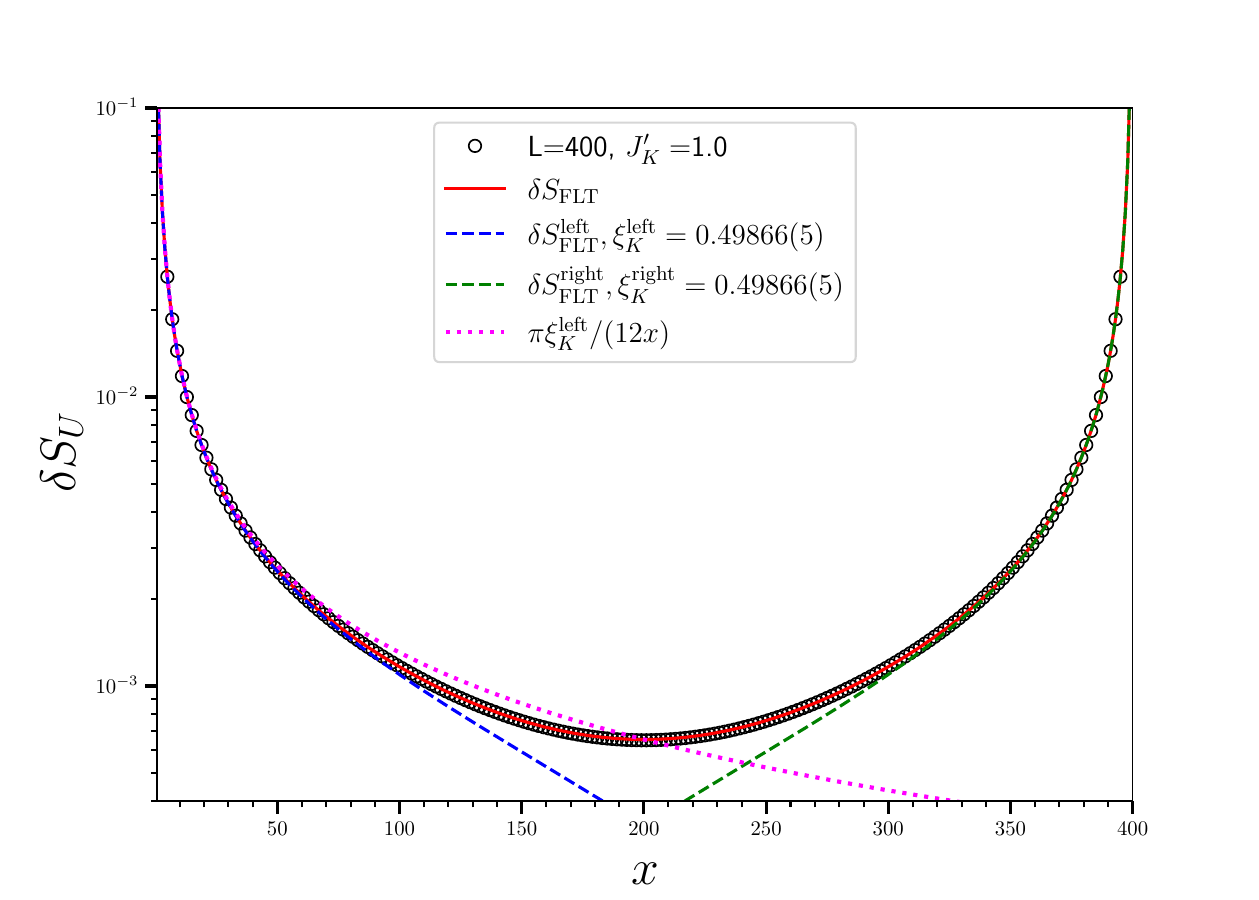}
\caption{$\delta S_U=S_U^\mathrm{obc}-S_U^\mathrm{bulk}$ from DMRG calculations with $\epsilon=\myeps$ on an open uniform chain with $J'_K=1$ at $J_2^c=0.241167$ with $L=400$ (black circles). The blue and dark green dashed 
lines correspond to fits to $\dSL$, Eq.~(\ref{eq:dSleft}) and $\dSR$, Eq.~(\ref{eq:dSright}), respectively. The solid red line is a fit to the combined
FLT expression, $\dSFLT$, Eq.~(\ref{eq:dSFLT}). The dotted magenta line is the thermodynamic limit FLT result, Eq.~(\ref{eq:dSthermo}). In all cases $\xi_K^{\rm left}=\xi_K^{\rm right}=\xi_K(J^\prime_K=1)=0.49866(5)$ was used.
}
\label{fig:1CK_FLTJK1}
\end{figure}

We begin by considering the uniform open chain, with $J_K'$=1, performing DMRG calculations on a system with $L=400$. We first extract the uniform part of the entanglement entropy, $S_U^\mathrm{obc}$. This is done using a 9-point stencil as described
in \ref{app:7pt}. We then fit the results for $S_U^\mathrm{obc}$ directly to Eq.~(\ref{eq:sb}). With only $\ln g+s_1/2$ and $\xi_K$ as the two free parameters, we obtain an excellent fit, see solid red line in Fig.~(\ref{fig:1CK_FLTJK1}). 
We explicitly determine:
\begin{equation}
\ln g + \frac{s_1}{2} = 0.198585(1)
\label{eq:s1lng}
\end{equation}
With the previously determined value for $s_1=\sone$ we can now determine $\ln g$ finding:
\begin{equation}
\ln g = 0.198585(1)-\sone/2 = -0.173287(4),
\end{equation}
in excellent agreement with the theoretical result, $\ln g = \ln 2^{-\frac{1}{4}}=-0.17328679$~\cite{Affleck1998,Zhou2006}. It is important to note that the precision of the numerical result
crucially relies on the detailed understanding of the correction term Eq.~(\ref{eq:dSFLT}). A similar analysis at a $J_2<J_2^c$ is complicated by additional corrections arising from the bulk marginal operator, rendering a high precision 
determination of $\ln g$ very challenging.
\begin{figure}[!ht]
\hfill\includegraphics[width=14cm,clip]{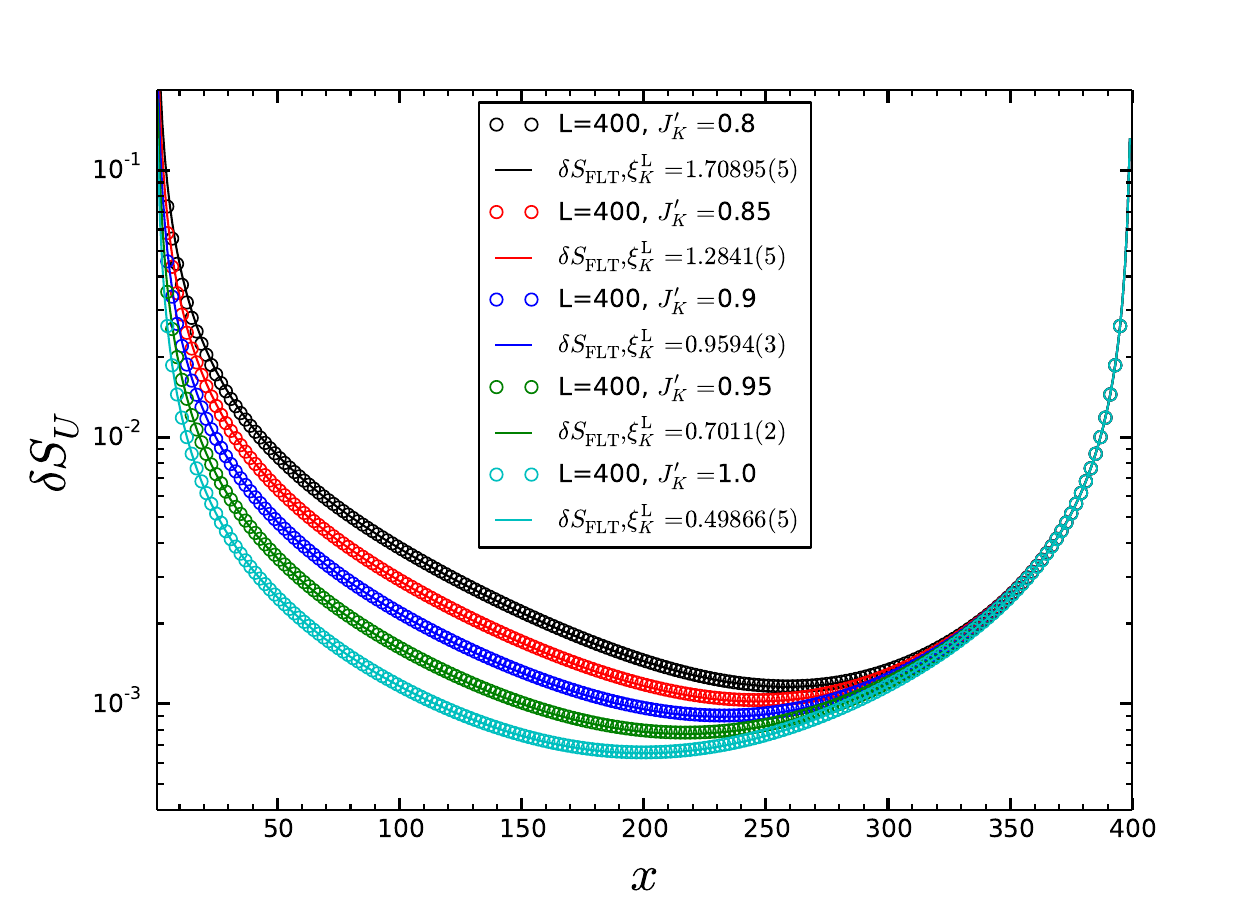}
\caption{$\delta S_U=S_U^\mathrm{obc}-S_U^\mathrm{bulk}$ from DMRG calculations with $\epsilon=\myeps$ on an open chain at $J_2^c=0.241167$ with $L=400$ and leftmost coupling $J^\prime_K=1,0.95,0.9,0.85,0.8$. 
The solid lines are fits to $\dSFLT=\dSL+\dSR$, Eq.~(\ref{eq:sb}) with $\xi^L_K(J^\prime_K)=0.49866(5)$, $0.7011(2)$, 
$0.9594(3)$,
$1.2841(5)$ and
$1.7090(5)$, respectively. In all cases with $\xi^R_K(J^\prime_K=1)=0.49866(5)$.
}
\label{fig:1CK_FLT2}
\end{figure}

In order to focus solely on the contribution arising from the two ends of the open chain we define:
\begin{equation}
\delta S_U=S_U^\mathrm{obc}-S_U^\mathrm{bulk},
\end{equation}
with $S_U^\mathrm{bulk}$ given in Eq.~(\ref{eq:Sbulk}). Our results for $\delta S_U$ for the uniform open chain with $J'_K=1$ are shown in Fig.~\ref{fig:1CK_FLTJK1}.
An excellent agreement between the numerical results (black circles) and the fit to Eq.~(\ref{eq:dSFLT}) (solid red line) is evident. Since the analytical expression is not valid
for $x\ll \xi_K$ we fit only data for $x>10 \xi_K$. 
With the above validation of the $\ln g$ term and the value of $s_1$ from the results with PBC, Eq.~(\ref{eq:s1}), we fix these parameters
and perform a one parameter fit in $\xi_K$ determining:
\begin{equation}
\xi_K(J^\prime_K=1)=0.49866(5).\ \ (J_2=J_2^c)
\end{equation}
We  note, that a similar analysis at $J_2=0$ yields a significantly larger estimate of $\xi_K$.
For completeness, the separate contributions from $\dSL$, Eq.~(\ref{eq:dSleft}) and $\dSR$, Eq.~(\ref{eq:dSright}) are shown as the dashed blue and green lines, respectively, in Fig.~\ref{fig:1CK_FLTJK1}. 
For comparison, we also show the thermodynamic limit FLT result, Eq.~(\ref{eq:dSthermo}) (dotted magenta line) which agrees very convincingly with the numerical results at small $x$
although the finite $L$ corrections, accounted for in Eq.~(\ref{eq:dSFLT}), are visible for $x>50$.

It is now possible to obtain very precise estimates for $\xi_K(J^\prime_K)$ as a function of $J^\prime_K$ by performing numerical calculations with 
a range of $J^\prime_K$ at the left end of the chain, (see Eq.~(\ref{eq:H1CK}), thereby varying $\xi_K^L$. With the known values of $\xi_K^R=0.49866(5)$ as well as $\ln g + \frac{s_1}{2}$, Eq.~(\ref{eq:s1lng}),
we then perform a one parameter fit of $\delta S_U$ to the analytical FLT form $\dSFLT=\dSL+\dSR$, Eq.~(\ref{eq:sb}). Our results are shown in Fig.~\ref{fig:1CK_FLT2} for $J^\prime_K=1,0.95,0.9,0.85,0.8$
with the solid lines representing the fitted form. An excellent agreement between the numerical data and analytical expression can be seen. Again, we restrict the fits to $10\xi_K < x$ to be sure we are in the FLT regime, Eq.~(\ref{eq:FLTreg}). We typically use $L=400$ and since we also must require
that $10\xi_K\ll L/2$ we stop the calculations at $J_K^\prime=0.5$ where $\xi_K^L$ reaches the value of 11.824(4). It would of course be relatively straight forward to repeat the calculations with $L>400$, however, since
$\xi_K\sim \exp(a/J_K^\prime)$ it quickly becomes unfeasible to extract $\xi_K(J_K^\prime)$ for smaller $J_K^\prime$ by fitting numerical results to Eq.~(\ref{eq:sb}).

\begin{table}[tbp]
\centering
\begin{tabular}{|c|l|c|l|c|l|c|l|c|l|}
\hline\hline
   $J'_K$ & $\xi_K$ &     $J'_K$ & $\xi_K$ &    $J'_K$ & $\xi_K$ &    $J'_K$ & $\xi_K$ &    $J'_K$ & $\xi_K$ \\
\hline
1.00 &  0.49866(5) &  0.90 &  0.9594(3) & 0.80 & 1.7293(5) & 0.70 & 3.066(2) &  0.60 & 5.708(2) \\
0.99 &  0.5361(1)  &  0.89 &  1.0308(3) & 0.79 & 1.8296(7) & 0.69 & 3.253(2) &  0.59 & 6.112(2) \\
0.98 &  0.5749(1)  &  0.88 &  1.0936(4) & 0.78 & 1.9396(8) & 0.68 & 3.456(2) &  0.58 & 6.540(2) \\
0.97 &  0.6152(1)  &  0.87 &  1.1595(4) & 0.77 & 2.0524(8) & 0.67 & 3.676(3) &  0.57 & 7.006(2) \\
0.96 &  0.6572(2)  &  0.86 &  1.2289(5) & 0.76 & 2.1720(9) & 0.66 & 3.906(1) &  0.56 & 7.524(2) \\
0.95 &  0.7011(2)  &  0.85 &  1.3018(5) & 0.75 & 2.2989(9)& 0.65 & 4.152(2)  &  0.55 & 8.082(2) \\
0.94 &  0.7513(2)  &  0.84 &  1.3785(5) & 0.74 & 2.4337(9) & 0.64 & 4.418(2) &  0.54 & 8.693(2) \\
0.93 &  0.7999(2)  &  0.83 &  1.4593(6) & 0.73 & 2.5770(9) & 0.63 & 4.704(2)&  0.53 & 9.365(3)\\
0.92 &  0.8507(3)  &  0.82 &  1.5445(6) & 0.72 & 2.730(1)  & 0.62 & 5.013(2)&  0.52 & 10.104(4)\\
0.91 &  0.9038(3)  &  0.81 &  1.6344(7) & 0.71 & 2.892(1)  & 0.61 & 5.346(2)&  0.51 & 10.920(5)\\

\hline\hline
\end{tabular}
\caption{$\xi _{K}(J_{K}^{\prime })$ for the 1CK  from DMRG calculations with $\epsilon=\myeps$ 
  on an open chain at $J_2^c=0.241167$ with $L=400$ determined by fitting the numerical results for
  $\delta S_U=S_U-S_A$ to
$\dSFLT=\dSL+\dSR$, Eq.~(\ref{eq:sb}). The error bar listed is the statistical error from the fit, which does not include systematic errors arising from corrections terms to Eq.~(\ref{eq:sb}) and other sources such as the range of $x$ used in the fit.
}
\label{tab:xiKFLT}
\end{table}
It is often very useful to have high precision numerical data for $\xi_K(J^\prime_K)$ available, and we have collected some of our results for $\xi_K$ in Table~\ref{tab:xiKFLT}. Again, we stress that these results 
are obtained at $J_2=J_2^c$ and similar calculations at $J_2<J_2^c$ and in particular at $J_2=0$ yield larger values for $\xi_K$.

\section{$\xi_K$ as a function of $1/J^\prime_K$}
\begin{figure}[!ht]
\hfill\includegraphics[width=14cm,clip]{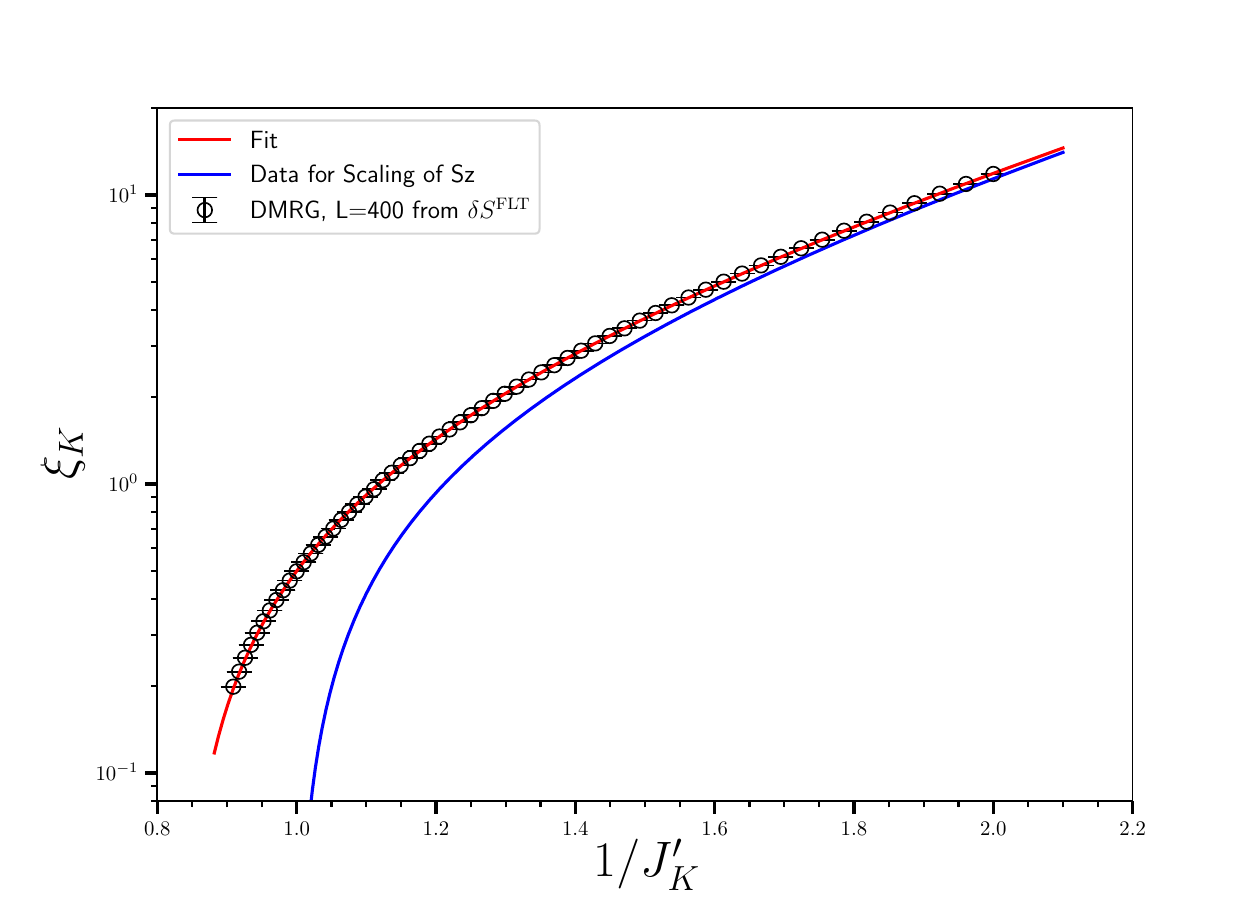}
\caption{Numerical data (open circles) for $\xi_K(J_K')$ as obtained from $\dSFLT$ (see Table~\ref{tab:xiKFLT}) fitted to the Eq.~(\ref{eq:xiKJK}) (solid red line)
  with the resulting fit given in Eq.~(\ref{eq:xiKfit}). The solid blue line indicates the $\xi_K(J_K')-\xi_K(J_K'=1)$ needed to scale $M^z$ and $N^z$.
}
\label{fig:1CK_xiK}
\end{figure}
Through perturbative calculations as detailed in \cite{laflorencie2008kondo}, $\xi_K$ can be related to the coupling $\lambda_K$ in Eq.~(\ref{eq:H1D}) as follows:
$
\xi_K=(\hbox{constant}/\sqrt{\lambda_K})\exp \left(1/\lambda_K\right)
\left[1+O(\lambda_K)\right].
$
Furthermore, through a careful determination of the spin velocity $v_s=1.174(1)$ and by studying end to end correlation functions, it was shown~\cite{laflorencie2008kondo} that in the limit $J_K^\prime\to 0$, at $J_2^c$, 
the following relation between $J_K^\prime$ and $\lambda_K$ holds 
\begin{equation}
J_K'= 1.381~\lambda_K.
\end{equation}
In terms of $J_K^\prime$ we therefore have
\begin{equation}
\xi_K=\frac{\hbox{constant}}{\sqrt{J_K^\prime}}\exp \left(1.381/J_K^\prime\right)
\left[1+O(J_K^\prime)\right].
\label{eq:xiKJK}
\end{equation}
Here, the $O(J_K^\prime)$ term represents a power series in $J_K^\prime$.
Note that, for $J_2=0$, the behavior is slightly different and instead one has $\xi_K\sim \exp(c/\sqrt{J_K^\prime})$~\cite{laflorencie2008kondo}. 
It is straightforward to verify that our results obey this relation
very precisely. As shown in Fig.~\ref{fig:1CK_xiK} we can fit the numerical data from Table~\ref{tab:xiKFLT} to Eq.~(\ref{eq:xiKJK}) obtaining
\begin{equation}
  \xi_K=\frac{1.002(1) }{\sqrt{J_K^\prime}}\exp \left( 1.381/J_K^\prime\right)\left(1-0.963(1)J_K^\prime+0.089(1)\left(J_K^\prime\right)^3\right).
\label{eq:xiKfit}
\end{equation}
In this fit the previously estimated constant $1.381$ is held fixed and as is clearly evident in Fig.~\ref{fig:1CK_xiK} at large $1/J_K^\prime$ the exponential divergence 
is clearly in agreement with this independently determined constant. The addition of the correction terms allows the form Eq.~(\ref{eq:xiKJK}) to fit the data even for 
relatively large $J_K^\prime$ even exceeding 1. This indicates that a critical value $J_K^{\prime c}$ exists at which the correction $\dSFLT$ vanishes. This coupling 
can be estimated to be:
\begin{equation}
J_K^{\prime c}=1.188(2).`
\label{eq:JKc}
\end{equation}

\section{Scaling of $\langle S^z(x)\rangle$ for Odd Length Chains}
\begin{figure}[!ht]
\hfill\includegraphics[width=14cm,clip]{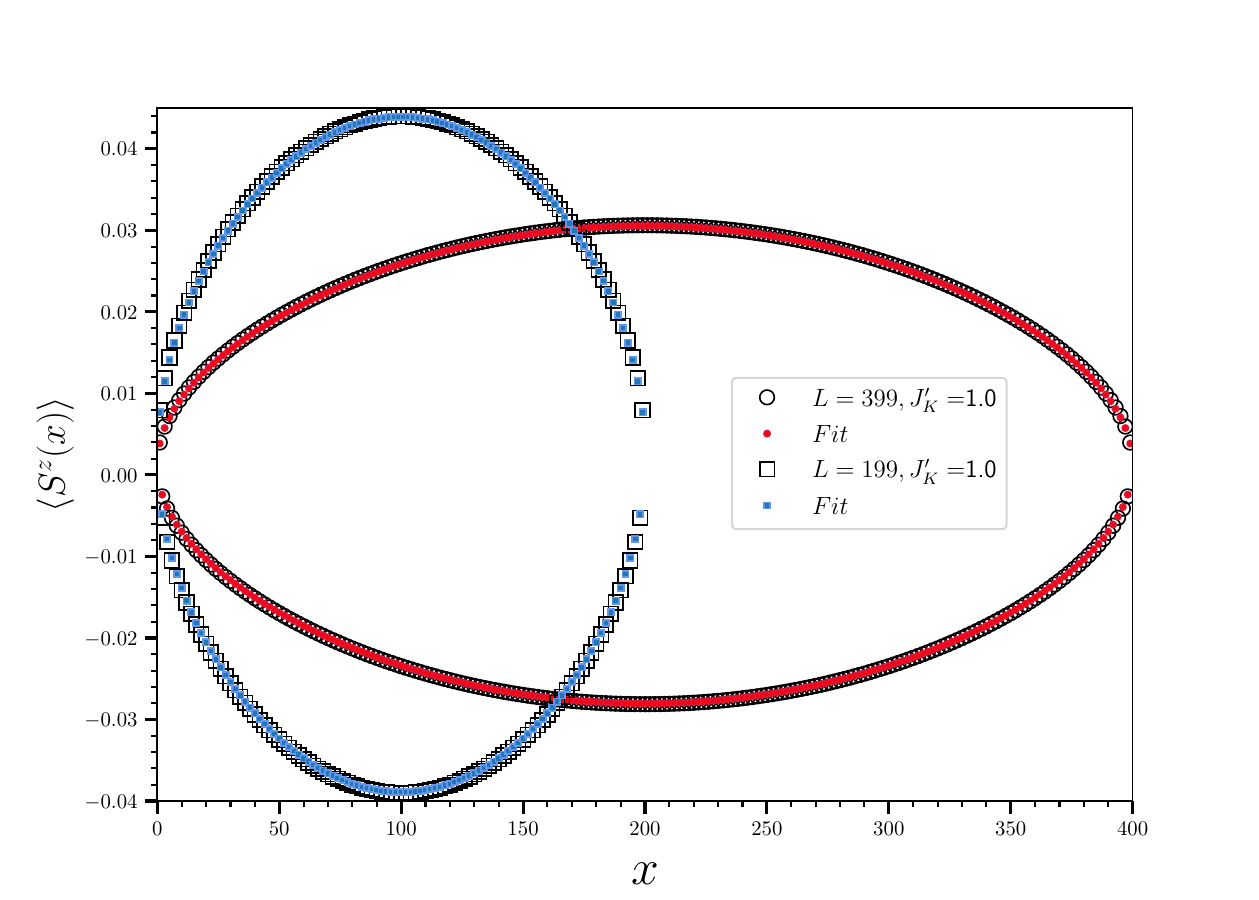}
\caption{$\langle S^z_i\rangle$ for a uniform ($J_K^\prime=1$) spin chain at $J_2^c$ of length $L=199$ (squares) and $399$ (circles). The solid points are fits
  to the form Eq.~(\ref{eq:szi}) including a constant, $1/(2L+3)$, representing the uniform part.
}
\label{fig:1CK_Szi}
\end{figure}
We now turn to an analysis of scaling of the magnetization $\langle S^z_i\rangle$ occurring in {\it odd} length systems and we focus
exclusively on the ground-state with $S^z_\mathrm{Tot}\equiv\sum_x S^z(x)$=1/2.
Since the system is antiferromagnetic and the length is odd, $\langle S^z(x)\rangle$ will have both a uniform and alternating part.
We write:
\begin{equation}
    \langle S^z(x)\rangle=M^z(x)+(-1)^{x+1} N^z(x),
\label{eq:szx}
\end{equation}
Our goal here is to develop a scaling form for the uniform and alternating part of the magnetization as the coupling to
the impurity spin, $J'_K$, is varied.
In the case of the uniform chain with ($J_K'=1$) at $J_2^c$, it was shown~\cite{eggert2002} that for the {\it alternating} part, $N^z$, of $\langle S^z(x)\rangle$ in the $S^z_\mathrm{Tot}=1/2$ subspace:
\begin{equation}
N^z(x)\simeq C\sqrt{\frac{\pi}{2L+3}\sin\frac{\pi x}{L+1}}.
\label{eq:szi}
\end{equation}
At small $x$ the alternating part $N^z(x)$ then increases in a characteristic $\sqrt{x}$ manner.
There is also a uniform part, $M^z$, which in the middle of the chain is well approximated by the constant $1/(2L+3)$. 
We note that for the calculations we consider here in the $S^z_\mathrm{Tot}=1/2$ subspace we must have $S^z_{Tot}=\sum_x S^z(x)=\sum_x (M^z(x)+(-1)^{x+1}N^z(x))=1/2$.
Our fits to this form are shown in Fig.~\ref{fig:1CK_Szi}
yielding excellent agreement with Eq.~(\ref{eq:szi}) allowing us to determine $C=0.46788(1)$. Here, the calculations are performed at $J_2^c$. At smaller values of $J_2$ corrections to
the form Eq.~(\ref{eq:szi}) occur~\cite{sanyal2011}.

\begin{figure}[!ht]
\hfill\includegraphics[width=14cm,clip]{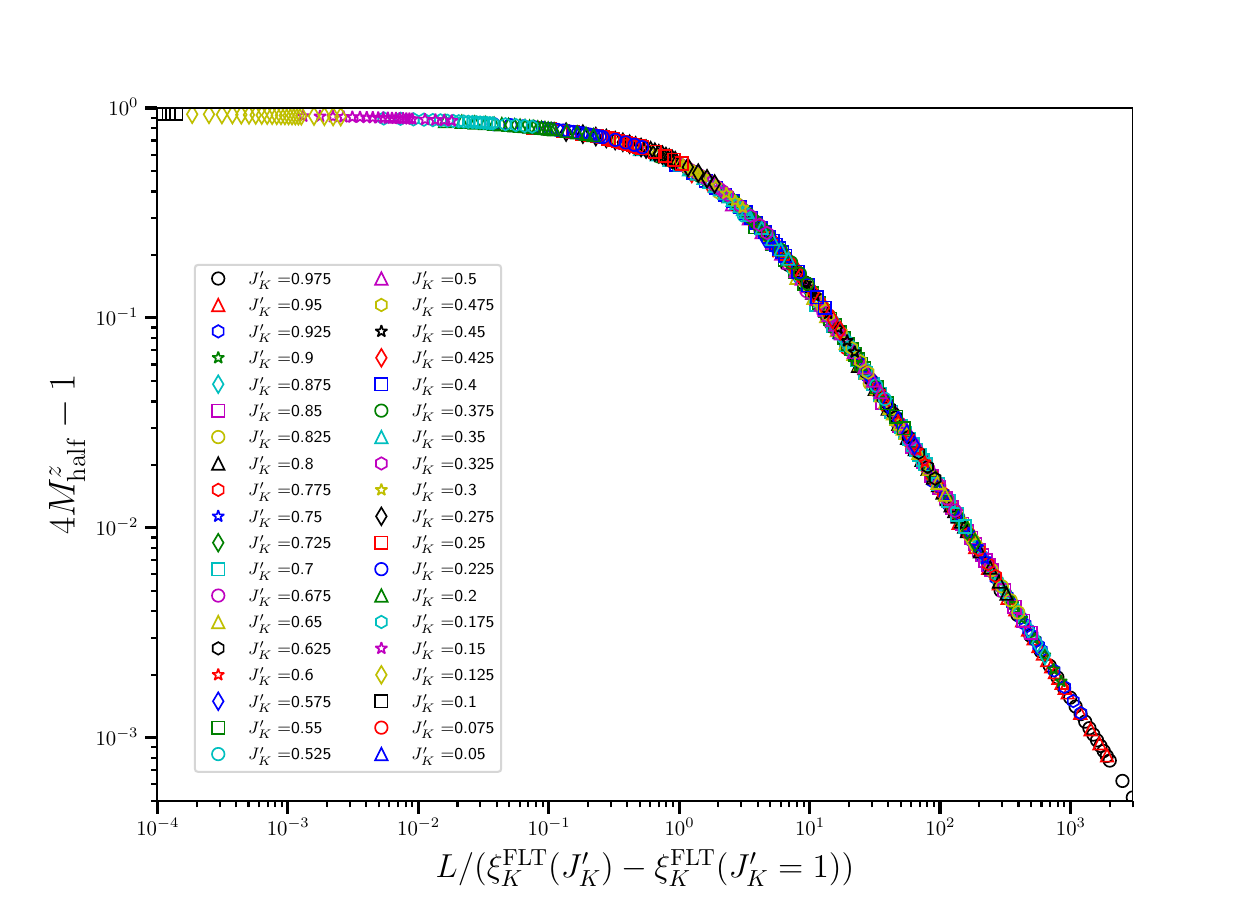}
\caption{The $4M^z_\mathrm{half} -1$ for a range of $J_K^\prime$ between $0$ and $1$, and 
$L= 29,\ldots,399$. 
Results are shown to scale when plotted versus $L/(\xi_K^\mathrm{FLT}-\xi_K^\mathrm{FLT}(J_K^\prime=1))$.
All calculations are at $J_2=J_2^c$ in the subspace with $S^z_\mathrm{Tot}=1/2$.
}
\label{fig:1CK_4HM1}
\end{figure}
At the isotropic point, $J_K^\prime=1$,
the on-site magnetization $\langle S^z(x)\rangle$ is also symmetric around the site $x=(L+1)/2$ which means that we can define the half-magnetization:
\begin{equation}
M^z_\mathrm{half}=\sum_{x=1}^{(L-1)/2} S^z(x) + \frac{1}{2} S^z(x=(L+1)/2).
\label{eq:H}
\end{equation}
Since $S^z_{Tot}=1/2$ and since $M^z_\mathrm{half}$ encompasses half of this magnetization when $J'_K=1$ it follows 
that $4M^z_\mathrm{half}-1= 0$ at $J_K^\prime =1$. Note that the impurity spin resides on site $x=1$. In the other limit of $J_K^\prime\to 0$ we instead find $4M^z_\mathrm{half}-1\to 1$ since the effectively decoupled impurity spin
at $x=1$ will carry all the magnetization so that $4M^z_\mathrm{half}=2$. However, for intermediate, $J_K^\prime$ we expect $4M^z_\mathrm{half}-1$ to only depend on the single variable $L/\xi_K$ and follow the scaling form:
\begin{equation}
4M^z_\mathrm{half}-1 \equiv h(L/\xi_K),
\label{eq:hscale}
\end{equation}
as $J_K^\prime$ is varied between $0$ and $1$. We want to demonstrate this scaling using the previously determined $\xi_K^\mathrm{FLT}$, however, for the quantity $4M^z_\mathrm{half}-1$ the strong coupling fixed point
where $\xi_K=0$ corresponds $J_K^\prime=1$ whereas for $\dSFLT$ $\xi_K=0$ occurs for $J_K^{\prime c}=1.188(2)$. In order to correct for this difference, we therefore need to scale the data for
$4M^z_\mathrm{half}-1$ using $\xi_K^\mathrm{FLT}-\xi_K^\mathrm{FLT}(J_K^\prime=1)$ instead of just $\xi_K^\mathrm{FLT}$. In Fig.~\ref{fig:1CK_4HM1} we show numerical results for $4M^z_\mathrm{half}-1$ displaying excellent scaling
with $L/(\xi_K^\mathrm{FLT}-\xi_K^\mathrm{FLT}(J_K^\prime=1))$ with $\xi_K^\mathrm{FLT}$ obtained from $\dSFLT$. (See Table~\ref{tab:xiKFLT}.)

\begin{figure}[!ht]
\hfill\includegraphics[width=16cm,clip]{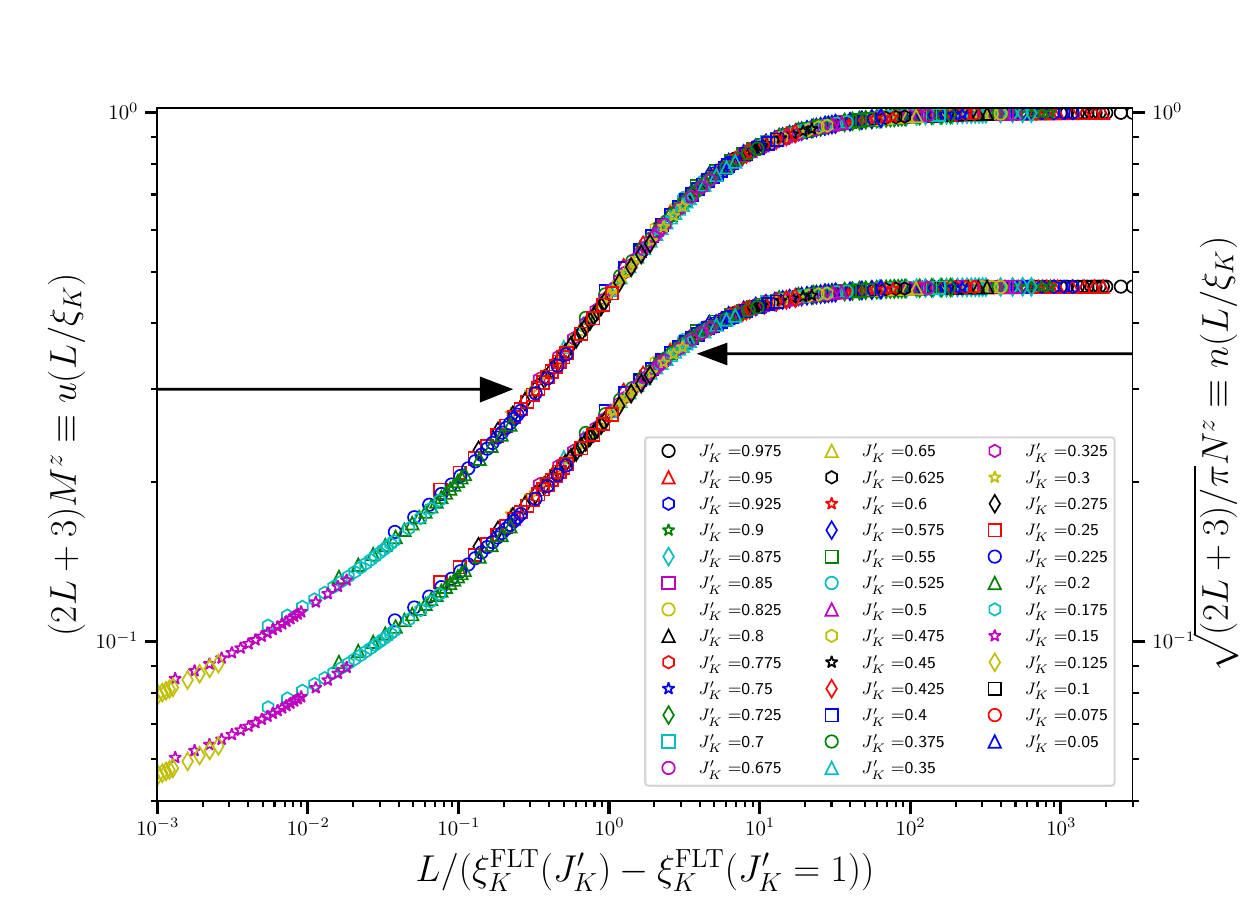}
\caption{Leftmost data set: The scaled uniform part of $\langle S^z(x=(L+1)/2)\rangle$ in the middle of the chain $(2L+3)M^{z}$, for a range of $J_K^\prime$ between $0$ and $1$ and 
$L= 29,\ldots,399$. Rightmost data set: The scaled alternating part of $\langle S^z(x=(L+1)/2)\rangle$ in the middle of the chain, $\sqrt{(2L+3)/\pi} N^z$, for the same range of $J_K^\prime$ and $L$.
The results clearly scale when plotted versus $L/(\xi_K^\mathrm{FLT}-\xi_K^\mathrm{FLT}(J_K^\prime=1))$.
All calculations are at $J_2=J_2^c$ in the subspace with $S^z_\mathrm{Tot}=1/2$.
}
\label{fig:1CK_NU2}
\end{figure}
Building on the relations, Eq.~(\ref{eq:szx}) and (\ref{eq:szi}), we can also construct scaling forms for $M^{z}(x)$ and $N^z(x)$.
Again we focus on the middle of the chain $x=(L+1)/2$ and extract the uniform and alternating part of $\langle S^z(x=(L+1)/2)\rangle$. We then expect to find:
\begin{equation}
(2L+3)M^{z}\equiv u(L/\xi_K) 
\label{eq:uscale}
\end{equation}
as well as
\begin{equation}
\sqrt{\frac{2L+3}{\pi}} N^z \equiv n(L/\xi_K),
\label{eq:nscale}
\end{equation}
with $u$ and $n$ scaling functions. We note that for the impurity spin itself and for the site coupled to it, weak scaling violations should
be observed~\cite{laflorencie2008kondo} due to the presence of an anomalous dimension for the corresponding operator~\cite{Barzykin1998}.
For both the prefactors of $2L+3$ for $M^z$ and $\sqrt{(2L+3)/\pi}$ for $N^z$, it is only the overall $L$ dependence that is known, with the precise additive constant determined numerically. As was the case for $M^z_\mathrm{half}$, the strong coupling fixed point where $\xi_K=0$ again corresponds to $J_K^\prime=1$ and we therefore
again use $\xi_K^\mathrm{FLT}-\xi_K^\mathrm{FLT}(J_K^\prime=1)$ instead of just $\xi_K^\mathrm{FLT}$. Our results are shown in Fig.~\ref{fig:1CK_NU2} where data for both
the scaled $M^{z}$ and $N^z$ are shown together. An excellent scaling of the data when plotted versus $L/(\xi_K^\mathrm{FLT}-\xi_K^\mathrm{FLT}(J_K^\prime=1))$ is evident. However,
in the limit $L\ll\xi_K$ deviations are clearly visible as one would expect. Similar deviations from scaling for $L\ll\xi_K$ are visible in Fig.~(\ref{fig:1CK_4HM1}) but are obscured by
the fact that $4M^z_\mathrm{half}-1$ approaches 1 in the $J_K^\prime\to 0$ limit.

\section{Discussion}
Using a spin chain representation of the one-channel Kondo model we have shown that scaling with the Kondo length scale $\xi_K$ is obeyed
to a high degree of precision and that reliable estimates of $\xi_K$ can be obtained. For the open chain it is clear that apart from the
usual term proportional to the central charge, the entanglement entropy contains two other terms, an alternating term $S_A$ but also an impurity
term arising from the boundary. We have also demonstrated that the same length scale, $\xi_K$, from our analysis of the entanglement entropy can be used to scale the magnetization present in a sub-system of the chain if a proper normalization is taken into account. The analysis of the magnetization is in some sense simpler since it directly yields the impurity contribution without involving a subtractive procedure.

\ack
The author acknowledges many valuable discussions with Ian Affleck,
and acknowledges the support of
the Natural Sciences and Engineering Research Council of Canada (NSERC) through Discovery
Grant No. RGPIN-2017-05759.
This research was enabled in part by support provided by SHARCNET (sharcnet.ca) and the Digital Research Alliance of Canada (alliancecan.ca).
Part of the numerical
calculations were performed using the ITensor library~\cite{itensor}.

\appendix

\section{Extracting the Uniform and Alternating Parts}
\label{app:7pt}
For the analysis of the data, it is necessary to extract the uniform and alternating part of a function defined on discrete points $i$. Focusing on a point $i=0$, we make the following assumption~\cite{sorensen2007QIE}:
\begin{equation}
f(i)=u(i)+(-1)^is(i).
\end{equation}
In the vicinity of the point $i=0$ we can then develop a 
7-point stencil by assuming that $u$ and $s$ are well approximated by slowly varying polynomials:
$u(i)\simeq a i^3+b i^2+c  i + d$ and $s(i)=e i^2+f i +g$,  for $i=0,\pm 1, \pm 2, \pm 3$, arriving
at the equations:
\begin{eqnarray}
f(i-3)&\simeq&-27a+9b-3c+d-(9e-3f+g)  \nonumber \\
f(i-2)&\simeq&-8a+4b-2c+d+(4e-2f+g)  \nonumber \\
f(i-1)&\simeq&-a+b-c+d-(e-f+g)  \nonumber \\
f(i)&\simeq&d+g  \nonumber \\
f(i+1)&\simeq&a+b+c+d-(e+f+g)  \nonumber \\
f(i+2)&\simeq&8a+4b+2c+d+(4e+2f+g)  \nonumber \\
f(i+3)&\simeq&27a+9b+3c+d-(9e+3f+g)  \nonumber \\
\end{eqnarray}
As can easily be verified,
the solutions for $u,s$ are then:
\begin{eqnarray}
u(i)&=& -\frac{1}{32}f(i-3) +\frac{9}{32}f(i-1)+\frac{1}{2}f(i)\nonumber\\
    & & -\frac{1}{32}f(i+3) +\frac{9}{32}f(i+1).\\
s(i)&=&f(i)-u(i)
\end{eqnarray}
In Ref.~\cite{sorensen2007QIE} a slight variation on these solutions were used.
The above equations have the advantage that they are explicitly invariant under $i-3\leftrightarrow i+3$, $i-2\leftrightarrow i+2$ and $i-1\leftrightarrow i+1$ as one would expect. Note that, as we move along the chain, the sign of $g$ will vary.
An equivalent 9-point stencil can be derived, yielding:
\begin{eqnarray}
u(i)&=&\frac{3}{256}f(i-4) -\frac{1}{32}f(i-3) -\frac{3}{64}f(i-2)+\frac{9}{32}f(i-1)+\frac{73}{128}f(i)\nonumber\\
     &+ &\frac{3}{256}f(i+4)
       -\frac{1}{32}f(i+3) 
       -\frac{3}{64}f(i+2)
      +\frac{9}{32}f(i+1).\\
s(i)&=&f(i)-u(i)
\end{eqnarray}

\newpage
\bibliographystyle{unsrt}
\bibliography{qie2ck}
\end{document}